\documentclass[12pt]{iopart}
\usepackage{graphicx}

\begin{document}

\title[Comparison between experiment and theory in Casimir 
force measurements]{Rigorous approach to the comparison 
between experiment and theory in Casimir force measurements
}
\author{
G~L~Klimchitskaya$^{1}$,
F~Chen$^2$, R~S~Decca$^{3}$,
E~Fischbach$^4$, D~E~Krause$^{5,4}$,
D~L\'{o}pez$^{6}$, U~Mohideen$^{2}$
and V~M~Mostepanenko$^{7}$}

\address{$^1$
North-West Technical University, Millionnaya St. 5,
St.Petersburg, Russia
}
\address{$^2$
Department of Physics, University of California, Riverside,
CA 92521, USA}
\address{$^3$
Department of Physics, Indiana University-Purdue University
Indianapolis, Indianapolis, IN 46202, USA}
\address{$^4$
Department of Physics, Purdue University,
West Lafayette, IN 47907, USA}
\address{$^5$
Department of Physics, Wabash College,
Crawfordsville, IN 47933, USA}
\address{$^6$ 
Bell Laboratories, Lucent Technologies,
Murray Hill, NJ 07974, USA}
\address{$^7$
Noncommercial Partnership  ``Scientific Instruments'', Moscow, Russia.}

\begin{abstract}
In most experiments on the Casimir force the
comparison between measurement data and theory
was done using the
concept of the root-mean-square deviation, a procedure
that has been criticized in literature. Here we propose
a special statistical analysis which should be performed
separately for the experimental data and for the results of
the theoretical computations. In so doing, the random,
systematic, and total experimental errors are found as functions 
of separation, taking into account the distribution laws for
each error at 95\% confidence. Independently, all
theoretical errors are combined to obtain the total theoretical 
error at the same confidence. Finally, the confidence interval
for the differences between theoretical and experimental
values is obtained as a function of separation. This
rigorous approach is applied to two recent
experiments on the Casimir effect.
\end{abstract}

\section{Introduction}

Today the Casimir effect is being actively investigated not only
theoretically but also experimentally. Historically the first
measurement of the Casimir force between metals was performed
in 1958 \cite{1} and confirmed the existence of the force
with an uncertainty of about 100\%. In the following decades
the experimental output was painfully low and only one
experiment with metal test bodies was made
\cite{2} (see Ref.~\cite{3} for a review). In the last few
years many measurements of the Casimir force have been performed
using torsion pendulums, atomic force microscopes, micromechanical
torsional oscillators, and other laboratory techniques [4--16].
Most authors (see Refs.~[1--14]) have used the concept
of the root-mean-square deviation between experiment and theory 
to quantify the precision of the measurements. However, for
strongly nonlinear quantities, such as the Casimir force which
changes rapidly with separation distance, this method is not
appropriate because it may lead to different results when applied
in different ranges of separations. This was emphasized in
Ref.~\cite{9} although no better method was suggested.

The present paper contains the comparison analysis of the
precision and accuracy in two recent experiments \cite{RD05,UM05}
using rigorous methods of mathematical statistics. The
distinctive feature of our approach is that both total
experimental and total theoretical errors are determined
independently of one another at some accepted confidence level.
Then, the absolute error of differences between calculated
and measured values of the physical quantity is found
at the same confidence as a function of separation, serving
as a measure of the precision in the comparison of experiment and theory.

\section{Determination of the experimental errors}

\subsection{Random errors}

In experiment \cite{RD05} the Casimir pressure 
between two Au coated parallel plates was determined dynamically 
by means of a microelectromechanical torsional oscillator
within the separation region from 160 to 750\,nm.
In experiment \cite{UM05} the Casimir force was
measured between a Si plate and a large Au coated sphere using
an atomic force microscope within the separations from
62.33 to 600.04\,nm. In our error analysis we use the
notation $\Pi(z)$ which denotes either the measured Casimir
pressure $P^{\rm exp}(z)$ or force $F^{\rm exp}(z)$ as a
function of separation $z$ between the test bodies.

Usually several sets of measurements, say $n$, are taken
within one separation region ($z_a,z_b$).
This is done in order to decrease the random error and to
narrow the confidence interval. In Ref.~\cite{RD05} $n=14$,
and in Ref.~\cite{UM05} $n=65$. Each set 
consists of pairs $[z_i,\Pi(z_i)]$ where
$1\leq i\leq i_{\max}=(288\div 293)$ in Ref.~\cite{RD05} and
$1\leq i\leq i_{\max}=3164$ in Ref.~\cite{UM05}.
All measurement data can be represented by pairs
$[z_{ij},\Pi(z_{ij})]$ where $1\leq j\leq n$.
Generally speaking, separations with fixed $i$
but different $j$ may be different (this was the case
in Ref.~\cite{RD05}). For such measurement results
it is reasonable to divide the entire separation range
($z_a,z_b$) into partial subintervals of length $2\Delta z$,
where $\Delta z$ is the absolute error in the measurement
of separations equal to 0.6\,nm and 0.8\,nm in 
Refs.~\cite{RD05,UM05}, respectively. In so doing, each
subinterval $k$ contains a group of $m_k$ points
$z_{ij}\equiv z_{kl}$, $1\leq l\leq m_k$ (in Ref.~\cite{RD05}
$m_k$ ranges from 3 to 13). Inside each subinterval all
points $z_{kl}$ can be considered as equivalent, because
within the interval of width $2\Delta z$ the value of
absolute separation is distributed uniformly. The mean and 
the variance of the mean of the physical quantity $\Pi$
for the subinterval $k$ are defined as
\begin{equation}
{\bar{\Pi}}_k=\frac{1}{m_k}
\sum\limits_{l=1}^{m_k}\Pi(z_{kl}),
\quad
s_{{\bar{\Pi}}_k}^2=\frac{1}{m_k(m_k-1)}
\sum\limits_{l=1}^{m_k}\left[\Pi(z_{kl})-{\bar{\Pi}}_k\right]^2.
\label{eq1}
\end{equation}

If all $z_{ij}=z_i$, i.e., the same in different sets of
measurement (as in Ref.~\cite{UM05}), the mean
and the variance of the mean at each point $z_i$ are
obtained more simply
\begin{equation}
{\bar{\Pi}}_i=\frac{1}{n}
\sum\limits_{j=1}^{n}\Pi(z_{ij}),
\quad
s_{{\bar{\Pi}}_i}^2=\frac{1}{n(n-1)}
\sum\limits_{j=1}^{n}\left[\Pi(z_{ij})-{\bar{\Pi}}_i\right]^2.
\label{eq2}
\end{equation}

Direct calculation shows that the mean values 
${\bar{\Pi}}_k,\,{\bar{\Pi}}_i$ are uniform, i.e., change
smoothly with the change of $k,\,i$. The variances of the mean,
$s_{{\bar{\Pi}}_k},\,s_{{\bar{\Pi}}_i}$, are, however, not
uniform. To smooth them,
we have used a special procedure developed in mathematical
statistics \cite{17,18}. At each separation $z_0$, in order
to find the uniform variance of a mean, we consider not only
one subinterval containing $z_0$ but also several neighboring
subintervals from both sides of $z_0$ (4 or 5
in Ref.~\cite{RD05}) or about 30 neighboring points in 
Ref.~\cite{UM05}. The number of neighboring subintervals 
or points is denoted by $N$. Then the smoothed variance of
the mean at a point $z_0$ is given by \cite{17,18}
\begin{equation}
s_{{\bar{\Pi}}}^2(z_0)=\max\left[N\sum\limits_{k=1}^{N}
\lambda_k^2s_{{\bar{\Pi}}_k}^2\right],
\label{eq3}
\end{equation}
\noindent
where $\lambda_k$ are the statistical weights.
The maximum in Eq.~(\ref{eq3}) is taken over two sets of
coefficients, $\lambda_k=1/N$, and
$\lambda_k=1/(c_k\sum_{i=1}^{N}c_i^{-1})$ where the constants 
$c_i$ are determined from 
$s_{{\bar{\Pi}}_1}^2:s_{{\bar{\Pi}}_2}^2:\ldots :s_{{\bar{\Pi}}_N}^2
=c_1:c_2:\ldots :c_N$. Note that max in Eq.~(\ref{eq3})
leads to the most conservative result, i.e., overestimates
the random error.
Finally, the confidence interval at a confidence probability
$\beta$ takes the form
\begin{equation}
\left[{\bar{\Pi}}(z_0)-\Delta^{\!\rm rand}{\Pi}(z_0),
{\bar{\Pi}}(z_0)+\Delta^{\!\rm rand}{\Pi}(z_0)\right],
\label{eq4}
\end{equation}
\noindent
where the random absolute error in the measurement of the
quantity $\Pi$ at a separation $z_0$ is given by
\begin{equation}
\Delta^{\!\rm rand}{\Pi}(z_0)=s_{{\bar{\Pi}}}(z_0)
t_{(1+\beta)/2}(\min m_k-1).
\label{eq5}
\end{equation}
\noindent
Here the value of $t_p(f)$ can be found in tables for
the Student's $t$-distribution.
For example, in the experiment\cite{UM05},  
 $\min m_k=n=65$. Thus, for $\beta=0.95$, we have
$t_p(f)=2.00$ and $\Delta^{\!\rm rand}F^{\rm exp}=3.0\,$pN
is independent of $z_0$.

The computational results for the relative random errors
$\delta^{\rm rand}{\Pi}=\Delta^{\!\rm rand}{\Pi}/|{\Pi}|$ 
in the experiments
\cite{RD05,UM05} at 95\% confidence are shown in columns labeled (a) in 
Table~1, as the functions of separation. As is seen from
column two in Table~1, in the experiment \cite{RD05} the
relative random error of the Casimir pressure measurements is equal
to 1.5\% at $z=160\,$nm, then it quickly decreases to 0.4\%
at $z=350\,$nm, and then increases with further increase of
separation. This is explained by the fact  that the absolute
random error in Eq.~(\ref{eq5}) takes a maximum value at the
shortest separation and monotonically decreases with the increase of
separation until $z=400\,$nm \cite{RD05}. At larger separations
$\Delta^{\!\rm rand}P^{\rm exp}$ is practically constant and the
increase of $\delta^{\rm rand}P^{\rm exp}$ is explained by solely in terms
of the decrease of the Casimir pressure magnitude. In the experiment
\cite{UM05} (column 8 in Table 1) the absolute random
error is only 0.78\% at the shortest separation $z=62.33\,$nm, and 
quickly increases with separation due to the decrease of the
Casimir force.

\subsection{Systematic errors}

\begin{table}
\caption{
Relative errors (\%) in experiments \cite{RD05,UM05}: 
random errors 
$\delta^{\rm rand}P^{\rm exp},\>\delta^{\rm rand}F^{\rm exp}$
(a); systematic errors
$\delta^{\rm syst}P^{\rm exp},\>\delta^{\rm syst}F^{\rm exp}$
(b);
total experimental errors
$\delta^{\rm tot}P^{\rm exp},\>\delta^{\rm tot}F^{\rm exp}$
(c);
theoretical errors 
$\delta_0P^{\rm th},\>\delta_0F^{\rm th}$ (d);
total theoretical errors
$\delta^{\rm tot}P^{\rm th},\>\delta^{\rm tot}F^{\rm th}$
(e). Columns labeled (f) contain $\Xi_P/|{\bar{P}}^{\rm exp}|$
and $\Xi_F/|{\bar{F}}^{\rm exp}|$ (see text).
}
\begin{indented}
\item[]\begin{tabular}{@{}cccccccccccccc}
\br
&\centre{6}{Experiment of Ref.~\cite{RD05}}
&\centre{6}{Experiment of Ref.~\cite{UM05}} \\
\ns&\crule{6}&{$\!\!$}&\crule{6}\\
$z\,$(nm)&(a)&(b)&(c)&(d)
&(e)&(f)&{$\!\!$}&(a)&(b)&(c)&(d)
&(e)&(f) \\
\mr
62.33&&&&&&&{$\!\!$}&0.78&0.31&0.87&0.55&3.5&4.0 \\
70&&&&&&&{$\!\!$}&1.1&0.42&1.2&0.56&3.2&3.7 \\
80&&&&&&&{$\!\!$}&1.6&0.60&1.7&0.56&2.8&3.7 \\
90&&&&&&&{$\!\!$}&2.1&0.84&2.4&0.56&2.6&3.9 \\
100&&&&&&&{$\!\!$}&2.9&1.1&3.2&0.56&2.4&4.4 \\
120&&&&&&&{$\!\!$}&4.7&1.8&5.3&0.56&2.0&6.2 \\
140&&&&&&&{$\!\!$}&7.3&2.8&8.1&0.57&1.8&9.1 \\
160&1.4&0.15&1.4&0.56&1.6&2.4&{$\!\!$}&10&4.1&12&0.58&1.6&13 \\
170&0.59&0.15&0.59&0.56&1.6&1.9&{$\!\!$}&12&4.9&14&0.58&1.6&15 \\
180&0.57&0.15&0.57&0.57&1.5&1.8&{$\!\!$}&15&5.7&16&0.58&1.5&18 \\
200&0.55&0.16&0.56&0.57&1.4&1.7&{$\!\!$}&20&7.7&22&0.59&1.4&23 \\
250&0.48&0.20&0.54&0.58&1.2&1.5&{$\!\!$}&37&14&41&0.61&1.3&42 \\
300&0.44&0.31&0.59&0.59&1.1&1.4&{$\!\!$}&62&24&69&0.64&1.2&70 \\
350&0.40&0.50&0.72&0.61&1.0&1.4&{$\!\!$}&96&37&107&0.67&1.1&108 \\
400&0.56&0.80&1.1&0.62&0.98&1.6&{$\!\!$}&&&&&& \\
500&1.3&1.80&2.5&0.66&0.91&2.9&{$\!\!$}&&&&&& \\
600&2.9&3.80&5.4&0.70&0.88&5.4&{$\!\!$}&&&&&& \\
\br
\end{tabular}
\end{indented}
\end{table}
In each of the experiments \cite{RD05,UM05} there are 
several absolute systematic errors 
$\Delta_i^{\!\rm syst}{\Pi}(z)$
and respective relative systematic errors
$\delta_i^{\rm syst}{\Pi}(z)=
\Delta_i^{\!\rm syst}{\Pi}(z)/|{\Pi}(z)|$
where $1\leq i\leq J$. Systematic errors are the random
quantities characterized by a uniform distribution.
Because of this, the total systematic error 
is \cite{20}
\begin{equation}
\delta^{\rm syst}{\Pi}(z)=\min\left\{\sum\limits_{i=1}^{J}
\delta_i^{\rm syst}{\Pi}(z),{\ }
k_{\beta}^{(J)}\sqrt{\sum\limits_{i=1}^{J}
\left[\delta_i^{\rm syst}{\Pi}(z)\right]^2}\right\},
\label{eq6}
\end{equation}
\noindent
where $\beta$ is the confidence probability, and 
$k_{\beta}^{(J)}$ is a tabulated coefficient \cite{20}. The same 
rule is also valid for the absolute systematic errors.

In the experiment \cite{RD05} there are $J=2$ main
systematic errors:
\begin{equation}
\delta_1^{\rm syst}P^{\rm exp}=\delta R=\frac{\Delta R}{R},
\quad
\delta_2^{\rm syst}P^{\rm exp}(z)=\delta (\omega_r-\omega_0)=
\frac{\Delta \omega_r}{|\omega_r-\omega_0|},
\label{eq7}
\end{equation}
\noindent
where $R=(148.7\pm 0.2)\,\mu$m is the sphere radius, 
$\omega_r$ and $\omega_0$ are the resonant and natural angular
frequencies of the oscillator, respectively (the former is
separation-dependent). $\omega_0=2\pi\times 702.92\,$Hz
was determined so precisely that its error does not contribute
to the results, and the error of the resonant frequency is
$\Delta\omega_r=2\pi\times 6\,$mHz. Using the value
$k_{0.95}^{(2)}=1.10$ and utilizing Eq.~(\ref{eq6})
one obtains the total systematic errors given in column 3
[labeled (b)] in Table 1.

The experiment \cite{UM05} contains the following $J=4$
systematic errors:
$\Delta_1^{\!{\rm syst}}F^{\rm exp}\approx 0.82\,$pN
due to force calibration; 
$\Delta_2^{\!{\rm syst}}F^{\rm exp}\approx 0.55\,$pN due to
noise when the calibration voltage is applied to the cantilever;
$\Delta_3^{\!{\rm syst}}F^{\rm exp}\approx 0.31\,$pN due to
the instrumental sensitivity; and
$\Delta_4^{\!{\rm syst}}F^{\rm exp}\approx 0.12\,$pN due to
the restrictions on computer resolution of data.
Combining these errors using the analog of Eq.~(\ref{eq6})
with $k_{0.95}^{(4)}=1.12$,
we obtain $\Delta^{\!{\rm syst}}F^{\rm exp}=1.17\,$pN.
The respective relative errors
$\delta^{{\rm syst}}F^{\rm exp}=
\Delta^{\!{\rm syst}}F^{\rm exp}/|F^{\rm exp}|$ are shown in
column 9 in Table 1. Comparing columns labeled (b) in
Table 1, we conclude that in both experiments the relative
systematic error increases as the separation increases. The
magnitudes of the systematic errors are smaller in
the experiment of Ref.~\cite{RD05}.

\subsection{Total experimental error}

To find the total experimental error in the measurements of 
$\Pi(z)$, one should combine the random and systematic errors 
obtained above which are described by a normal (or
Student) distribution and a combination of uniform distributions, 
respectively.  To be very conservative, we assume that
the systematic error is described by a uniform distribution
(other assumptions lead to smaller total error). Different
methods for combining random and systematic errors are
described in Ref.\cite{20}. Here we use one based
on the value of the quantity 
$r(z)=\Delta^{\!\rm syst}{\Pi}(z)/s_{\bar{\Pi}}(z)$. According
to this method, at all $z$ where $r(z)<0.8$ the contribution
from the systematic error is negligible and
$\Delta^{\!\rm tot}{\Pi}(z)=\Delta^{\!\rm rand}{\Pi}(z)$ at
95\% confidence. If $r(z)>8$ is valid, the random error is
negligible and at 95\% confidence
$\Delta^{\!\rm tot}{\Pi}(z)=\Delta^{\!\rm syst}{\Pi}(z)$.
In the separation region where $0.8\leq r(z)\leq 8$, the
combination of errors is performed using the rule
\begin{equation}
\Delta^{\!\rm tot}{\Pi}(z)=q_{\beta}(r)\left[
\Delta^{\!\rm rand}{\Pi}(z)+\Delta^{\!\rm syst}{\Pi}(z)\right],
\label{eq8}
\end{equation}
\noindent
where the coefficient $q_{\beta}(r)$ with $\beta=0.95$ varies
between 0.71 and 0.81. Being conservative, here we use
$q_{\beta}(r)=0.8$ in all calculations.

Table 1 [columns 4 and 10 labeled (c)] contain the total
experimental error of the Casimir pressure and force
measurements in the experiments
\cite{RD05,UM05}, respectively. As seen in
column 4 of Table 1, in the experiment \cite{RD05}
at $z=160\,$nm the total experimental error is equal to
1.4\%, but in a wide separation range from 170 to 300\,nm,
it is practically flat and within the range from 0.54
to 0.59\%. Even at $z=600\,$nm it is equal to only 5.4\%.
In the experiment \cite{UM05} (column 10 in Table 1)
the smallest total experimental error of 0.87\% is
achieved  $z=62.33\,$nm and
increases up to 5.3\% at $z=120\,$nm. This is mainly
due to the large contribution of the random errors.

\section{Determination of the theoretical errors}


The theoretical values of $\Pi(z)$ (both the pressure
and force) are computed using the Lifshitz formula (see, e.g., 
Ref.~\cite{3}) which takes into account the effects of finite
conductivity and nonzero temperature.
The Lifshitz formula contains the reflection coefficients
at imaginary Matsubara frequencies.
At zero Matsubara frequency these coefficients
are expressed in terms of the
Drude dielectric function (the Drude model approach \cite{21,22})
or in terms of the Leontovich surface impedance (the impedance 
approach \cite{23,24}). At nonzero Matsubara frequencies both
approaches use the tabulated optical data extrapolated to low
frequencies by the imaginary part of the Drude dielectric
function. In Refs.~\cite{25,26} the reflection coefficients
at all Matsubara frequencies were expressed using the free
electron plasma model (the plasma model approach).

One error in the theoretical computation arises from sample to
sample variations of the optical data for the complex index
of refraction. Usually these data are not measured in each
individual experiment, but are taken from Tables.
In Ref.~\cite{14} it was shown that variation of the
optical data for typical samples leads to a relative theoretical
error $\delta_1{\Pi}^{\rm th}(z)$ in the computed Casimir
pressure or force that is no larger than 0.5\%. Being conservative,
we set $\delta_1{\Pi}^{\rm th}(z)=0.5$\% at all separations. 
Strictly speaking, there may occur rare samples with
up to 2\% deviations in the Casimir pressure or force at
short separations.
If this happens, the theoretical values come into
conflict with the experimental data. Such deviations must
be considered not as an error (they can only diminish the
magnitudes of the pressure or force) but as a correction.
The validity of the hypothesis on the presence  of such types
of corrections can be easily verified statistically.

Another theoretical error is caused by the 
use of the proximity force theorem \cite{26a}.
(This is the name given by the authors of Ref.~\cite{26a};
some other authors, e.g. in Ref.~\cite{27}, prefer to use
the name ``proximity force approximation" to underline the
approximate character of the equality proposed in
Ref.~\cite{26a}.) In the experiment  
\cite{RD05} it is applied to express the effective
Casimir pressure between two parallel plates through the
derivative of the force acting between a sphere and a
plate. In the experiment \cite{UM05} the basic 
result for the force is obtained using the proximity
force theorem. The upper limit of error introduced by this is
$\delta_2{\Pi}^{\rm th}(z)=z/R$ \cite{3}
(see also Refs.~\cite{27,28} where the same estimation was
confirmed for the case of a massless scalar field).

Both errors $\delta_i{\Pi}^{\rm th}$ are described
by a uniform distribution and in this sense can be likened 
to systematic errors. They are combined by using 
Eq.~(\ref{eq6}) with $J=2$ leading to the values 
 $\delta_0{\Pi}^{\rm th}$ presented in columns 5 and 11 in
Table 1 [labeled (d)] for the experiments  
\cite{RD05,UM05}, respectively. As is seen from
these columns, the errors
$\delta_0{\Pi}^{\rm th}(z)$ 
depend only slightly on separation and take 
similar values between 0.55 and 0.70\%.

In addition to the major theoretical errors 
$\delta_i{\Pi}^{\rm th}(z)$, there exist other uncertainties
in calculations which are 
not taken into account in the Lifshitz formula. Some of them
were shown to be negligibly small (like the contributions
from patch potentials, nonlocal effects and finite sizes of the
plates \cite{14,RD05}). As to the contribution from the
surface roughness, it was calculated using the atomic force
microscope images of the interacting surfaces and taken into
account as a correction [14--16]. This is why 
these factors do not contribute to the balance of
theoretical errors.


There is one more error which can be considered 
together with the theoretical errors if one is going to
compare the experimental and theoretical values of
$\Pi(z)$ \cite{RD05,UM05}. This arises from
the fact that $z$ is determined experimentally with
an error $\Delta z$ (see Sec.~2.1), and this error results 
in the additional uncertainties $\delta_3{\Pi}^{\rm th}(z)$
in computations. Bearing in mind the leading theoretical
dependences of the pressure and force on separation, we
obtain $\delta_3P^{\rm th}(z)=4\Delta z/z$ in 
Ref.~\cite{RD05} and 
$\delta_3F^{\rm th}(z)=\Delta R/R+3\Delta z/z$ in
Ref.~\cite{UM05}. Taking into account that the combined
random quantity $\delta_0{\Pi}^{\rm th}(z)$
may be distributed nonuniformly, we
combine it with $\delta_3{\Pi}^{\rm th}(z)$ using
Eq.~(\ref{eq8}) and obtain the total theoretical
error $\delta^{\rm tot}{\Pi}^{\rm th}(z)$ at
95\% confidence. The values of
$\delta^{\rm tot}{\Pi}^{\rm th}(z)$ are presented in
columns 6 and 12 in Table 1 [labeled (e)] for the
experiments \cite{RD05,UM05}, respectively.
For both experiments they monotonically
decrease with separation and take the largest values 
at the shortest separation. The significant increase of the
total theoretical error in columns labeled (e) compared to
those labeled (d) is due to the additional error
$\delta_3{\Pi}^{\rm th}(z)$.

\section{Comparison between experiment and theory}

\subsection{Measure of agreement between 
experiment and theory}

In Secs.~2.3 and 3.2 we have obtained the total
experimental and theoretical errors at 95\% confidence
for both the Casimir pressure and force. Now we 
consider the new random quantity 
$P^{\rm th}(z)-P^{\rm exp}(z)$ [or 
$F^{\rm th}(z)-F^{\rm exp}(z)$] and determine the absolute
error of this quantity, $\Xi_{P,F}(z)$, at 95\% confidence using 
the composition rule (\ref{eq6}) with $J=2$
\begin{equation}
\fl
\Xi_P(z)=\min\left\{\Delta^{\!\rm tot}P^{\rm th}(z)+
\Delta^{\!\rm tot}P^{\rm exp}(z),1.1\sqrt{\left[
\Delta^{\!\rm tot}P^{\rm th}(z)\right]^2+\left[
\Delta^{\!\rm tot}P^{\rm exp}(z)\right]^2}\right\}
\label{eq9}
\end{equation}
\noindent
(the same equation is valid for the force).
Note that in Eq.~(\ref{eq9}) the conservative value
of $k_{0.95}^{(2)}=1.1$ is used as for two uniform 
distributions (otherwise it would be smaller).

The confidence interval for the quantity
$P^{\rm th}(z)-P^{\rm exp}(z)$ at 95\% confidence is given
by $[-\Xi_P(z),\Xi_P(z)]$ and the mean values 
$\langle P^{\rm th}(z)-P^{\rm exp}(z)\rangle$ or 
$\langle F^{\rm th}(z)-F^{\rm exp}(z)\rangle$
must belong to this interval or its analog for the force
with a 95\% probability. The values of
$\Xi_P(z)/|{\bar{P}}^{\rm exp}|$ and
$\Xi_F(z)/|{\bar{F}}^{\rm exp}|$ are given in columns 7 and
13 in Table 1 [labeled (f)] for the experiments 
\cite{RD05,UM05}, respectively. They characterize the
sensitivity of the experiments \cite{RD05,UM05}
to the differences between theory and experiment at 95\%
confidence. For example, in Ref.~\cite{RD05} theory is in 
agreement with experiment at a separation $z=400\,$nm
if $|P^{\rm th}(z)-{\bar{P}}^{\rm exp}(z)|$ does not exceed 1.6\% 
 of $|{\bar{P}}^{\rm exp}(z)|$.

\subsection{Comparison of experiment and theory in the determination
of the Casimir pressure between Au plates in Ref.~\cite{RD05}}

\begin{table}
\caption{
Comparison of the experiments of Refs.~\cite{RD05,UM05} 
with theory. Columns (a) contain the absolute errors
$\Xi_{P,F}(z)\,$ of the pressure (mPa) and force (pN) differences
at 95\% confidence. Column (b) contains the same quantity
for the pressure at 99\% confidence. Other columns contain
the mean values 
$\langle P^{\rm th}(z)-P^{\rm exp}(z)\rangle$ in mPa [and also 
$\langle F^{\rm th}(z)-F^{\rm exp}(z)\rangle$ in pN for the last
column labeled (e)] computed using four different approaches: the
impedance (c) and the plasma model (d) approach at
$T=300\,$K; the optical data in the Lifshitz formula
at $T=0$ (e); the Drude model approach at $T=300\,$K (f).
}
\begin{indented}
\item[]\begin{tabular}{@{}cccccccccr}
\br
&\centre{6}{Experiment of Ref.~\cite{RD05}}&
&\centre{2}{Experiment of Ref.~\cite{UM05}} \\
\ns&\crule{6}&&\crule{2}\\
$z\,$(nm)&(a)&(b)&(c)&(d)
&(e)&(f)&&(a)&(e) \\
\mr
62.33&&&&&&&&15.2&--0.5 \\
70&&&&&&&&10.4&3.0 \\
80&&&&&&&&7.1&3.6 \\
90&&&&&&&&5.4&1.0 \\
100&&&&&&&&4.5&2.0 \\
120&&&&&&&&3.9&--0.15 \\
140&&&&&&&&3.8&0.02 \\
170 & 17.2&39.8&2.01&13.0&3.87&18.8&&3.7&--0.82\\
180&13.4&31.0&--0.74&7.54&1.24&14.4&&3.7&--0.48 \\
200&8.59&19.8&--1.21&5.3&0.63&11.0&&3.7&--0.31 \\
250&3.34&7.72&--0.31&1.3&0.93&7.09&&3.7&--0.84 \\
300&1.59&3.67&0.34&0.6&1.12&5.07&&3.7&0.46\\
350&0.89&2.06&0.38&0.39&0.80&3.58&&3.7&0.27\\
400&0.63&1.46&0.28&0.20&0.68&2.59&&&\\
500&0.49&1.13&0.11&0.05&0.32&1.37&&&\\
600&0.46&1.06&0.08&0.04&0.17&0.82&&&\\
700&0.46&1.06&0.02&--0.01&0.08&0.51&&&\\
\br
\end{tabular}
\end{indented}
\end{table}
Experiment \cite{RD05} is rather sensitive and can be
compared with different theoretical approaches to the calculation
of the Casimir pressure. The main results are presented in 
Table 2 where the second and third columns labeled (a), (b)
contain the half-width $\Xi_P(z)$ of the confidence interval
at 95 and 99\% confidence, respectively.
In columns 4--7 labeled (c), (d), (e) and (f) the results for
the mean differences
$\langle P^{\rm th}(z)-P^{\rm exp}(z)\rangle$ are
computed using the impedance \cite{RD05,23} and the plasma model
\cite{RD05,25,26} approach at $T=300\,$K, the optical data in
the Lifshitz formula at $T=0$ \cite{3}, and the Drude model
approach at $T=300\,$K \cite{21,22}, respectively. 
To avoid confusion, recall that in column (c) the zero-frequency
contribution to the Lifshitz formula is computed using the
Leontovich impedance in the region of infrared optics.
At all other Matsubara frequencies the impedance is obtained
using the tabulated optical data. Comparing
columns 4--6 and columns 2,3, we conclude that the impedance
approach, the plasma model approach and the Lifshitz formula
at $T=0$ are consistent with the measurement data. At the 
same time, by comparing columns 2,3 with column 7 we find that
the Drude model approach is excluded by experiment at 95\%
confidence within the separation range from 170 to 700\,nm,
and at 99\% confidence from 300 to 500\,nm. 
The physical reasons for the failure of the Drude model
approach and the advantages of the Leontovich impedance
are discussed in Refs.~\cite{23,24,29}.

\subsection{Comparison of experiment and theory in
measuring the Casimir force between an Au sphere and a 
Si plate in Ref.~\cite{UM05}}

Experiment \cite{UM05} is the first demonstration of
the Casimir force between a metal and a semiconductor performed
at shorter separations than in experiment \cite{RD05}.
For this reason it cannot be used to discriminate among different
theories. In column 8 in Table 2 labeled (a) the values of
$\Xi_F(z)$ for the force at 95\% confidence are given. Column 9
in Table 2 labeled (e) contains the values of 
$\langle F^{\rm th}(z)-F^{\rm exp}(z)\rangle$
computed using the Lifshitz formula at $T=0$ and tabulated 
optical data for Au and Si. The comparison of these columns
shows that the theory at $T=0$ is in a very good agreement
with experiment.

\section{Conclusions}

From the above, several conclusions can be reached:

--- A new method for data processing and comparing
theory with experiment for the Casimir effect has been
presented based on rigorous results of mathematical
statistics with no recourse to the previously used 
root-mean-square deviation;

--- The distinguishing feature of this method is the independent
determination of the total experimental and theoretical errors
and of the confidence interval for 
differences between calculated and measured values at a chosen 
confidence probability;

--- The developed method is conservative and guaranties against 
underestimation of errors and uncertainties. It was applied to
two recent experiments measuring the Casimir pressure and
force in different configurations;

--- We have demonstrated that the approaches based on the vanishing
contribution of the transverse electric mode at zero frequency
(e.g., the Drude model approach) are excluded by experiment at
99\% confidence, whereas the three traditional approaches to the
thermal Casimir force are consistent with experiment.

\ack

The work of {GLK}, {FC}, {UM} and {VMM} was 
supported by the NSF Grant No PHY0355092 and DOE Grant No
DE-FG02-04ER46131. {EF} was supported by DOE Grant No
DE-AC02-76ER071428.


\end{document}